\documentstyle[aps,prb,multicol,epsf]{revtex}

\newcommand{\figurewidth}{83mm}
\begin{document}

\title{The heat capacity of the restricted primitive model electrolyte}

\author{Erik Luijten\thanks{Electronic address: luijten@ipst.umd.edu}
        and
        Michael E. Fisher}
\address{Institute for Physical Science and Technology,
         University of Maryland, College Park, MD 20742}

\author{Athanassios Z. Panagiotopoulos}
\address{Department of Chemical Engineering, Princeton University,
         Princeton, NJ 08540}

\date{November 2, 2000}

\maketitle

\begin{abstract}
  The constant-volume heat capacity, $C_V(T,\rho)$, of the restricted primitive
  model (RPM) electrolyte is considered in the vicinity of its critical point.
  It is demonstrated that, despite claims, recent simulations for finite
  systems do not convincingly indicate the absence of a divergence in
  $C_V(T,\rho)$---which would point to non-Ising-type criticality.  The strong
  qualitative difference between $C_V$ for the RPM and for a Lennard-Jones
  fluid is shown to result from the low critical density of the former.  If one
  considers the theoretically preferable configurational heat-capacity {\em
  density}, $C_V/V$, the finite-size results for the two systems display
  qualitatively similar behavior on near-critical isotherms.
\end{abstract}

\begin{multicols*}{2}
%\newpage
  
The critical behavior of Coulombic systems continues to be subject to debate.
Whereas it is generally accepted that the critical behavior of the gas--liquid
transition in simple liquids belongs to the three-dimensional~(3D) Ising
universality class, the situation in ionic solutions is considerably more
obscure.  At sufficiently low temperatures, these solutions exhibit separation
into two phases with different density, driven primarily by the Coulombic
forces between the charged constituents. Experimentally, both classical (as one
might guess from the long-range character of the ionic forces) and Ising-type
critical behavior (as might be explained by the effects of Debye screening)
have been reported: see, e.g., Refs.~\onlinecite{fisher94,fisher96}.  Other
possible scenarios entail a crossover from classical to Ising-type behavior at
considerably smaller reduced temperatures than in simple fluids or even the
existence of a different type of criticality.\cite{fisher94,fisher96}

In view of the significance of electrolytes and ionic systems in many domains,
a clear understanding of their critical behavior is of interest.  It is,
therefore, disconcerting that even for the simplest model thought to capture
the salient features of such systems, namely, the {\em restricted primitive
model\/} (RPM), the universality class has not yet been established beyond
reasonable doubt. The RPM consists of a mixture of hard spheres of uniform
diameter~$\sigma$, half of which carry a charge~$+q$ and half a charge~$-q$.
Its critical behavior has been analyzed by both analytical and numerical means.
Analytically, a fairly satisfactory description of the critical region (except
for the nature of the criticality) has been obtained from Debye--H\"uckel
theory supplemented by Bjerrum's concept of ion pairing and allowance for the
solvation of dipolar--ion pairs in the ionic fluid.\cite{fisher93} However,
lack of a sufficiently adequate formulation at the mean-field level has
hindered the development of a renormalization-group treatment, see, e.g.,
Ref.~\onlinecite{moreira99}.  Furthermore, simulations have also encountered
serious difficulties, not only because of the long-range nature of the
interactions, but, in particular, because of the low value of the critical
temperature and the resulting presence of many strongly bound ion
pairs.\cite{caillol94,orkoulas94,shelley95,caillol96} The limited statistical
accuracy and range in system sizes that have been reliably accessed have
hampered detailed numerical analysis.

This note has been inspired by recent work by Valleau and Torrie
(VT),\cite{valleau98} who performed numerical simulations of the RPM using a
temperature-and-density-scaling Monte Carlo method.  Other
simulations\cite{caillol94,orkoulas94,shelley95,caillol96,yan99} focused mainly
on the coexistence curve below~$T_c$. Experimentally, observations of the
coexistence curve as $T \to T_c-$ have been revealing of universality class
(with $\beta_{\rm Ising} \simeq 0.326$ and $\beta_{\rm classical} =
\frac{1}{2}$) or of crossover behavior. In simulations, however, finite-size
effects preclude the estimation of the coexistence curve close to $T_c$:
Wilding and Bruce\cite{wilding92} have devised a finite-size scaling technique
for analyzing the corresponding Monte Carlo data which has led to fairly
precise and seemingly rather reliable estimates of the critical temperature,
$T_c$, and to reasonable estimates of the overall ionic critical density,
$\rho_c$.  However, their technique {\em presupposes\/} Ising-type criticality
and has not, therefore, provided any effective criteria for ruling out (or,
possibly, revealing) other types of criticality.

\begin{figure}
\leavevmode
\centering
\epsfxsize \figurewidth
\epsfbox{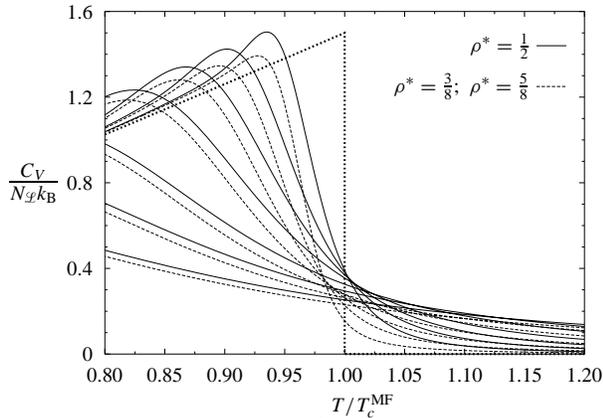}
\caption{The specific heat, $C_V/N_{\cal L}k_{\rm B}$, for a grand-canonical
lattice gas with identical interactions between all particle pairs, near the
mean-field critical temperature $T_c^{\rm MF}$.  The solid curves derive (in
order of increasing peak height) from finite systems of $N_{\cal L}=10, 20, 40,
100, 200, 400, 1000$ lattice sites, for a mean particle density $\rho = \rho_c
= \frac{1}{2}\rho_{\rm max}$.  The corresponding dashed curves pertain to
densities $\rho = \frac{3}{4}\rho_c$ and $\rho = \frac{5}{4}\rho_c$. The bold
dotted curve represents the thermodynamic limit, $N_{\cal L} \to \infty$, for
$\rho=\rho_c$. The maxima of the finite-size curves saturate at
approximately~$1.657$, rather than at the thermodynamic
maximum~$\frac{3}{2}$.\protect\cite{thesis}}
\label{fig:mf}
\end{figure}

By contrast, VT\cite{valleau98} focused on the heat capacity at constant
volume, $C_V(T,\rho)$, in the one-phase region both as a function of density,
$\rho$, near~$T_c$ and on approach to criticality from above. In a classical,
or van der Waals-type system $C_V$ remains finite as $T\to T_c+$ on the
critical isochore, $\rho=\rho_c$, whereas in an Ising-type system,
$C_V(T,\rho_c)$ diverges to infinity, albeit weakly with an exponent $\alpha
\simeq 0.109$. As one passes through $T_c$ from above in a classical system,
$C_V(T,\rho_c)$ undergoes a positive jump discontinuity and decreases smoothly
thereafter: see Fig.~\ref{fig:mf}; an Ising-type fluid exhibits a
$|T-T_c|^{-\alpha}$ singularity falling rapidly from infinity as $T$ decreases.
Accordingly, VT argued that an examination of $C_V(T,\rho)$ for the RPM for $T
\gtrsim T_c$ and, in particular, comparison with simulations of a
Lennard-Jones~(LJ) model fluid (for which Ising-type or close-to-Ising-type
behavior may be accepted) should provide an effective diagnostic of critical
behavior. On the basis of the simulations they undertook and presented, VT
concluded that little if any evidence of a rise in specific heat was present in
the RPM\@. This suggested that criticality in the RPM might be classical in
nature or, at least, characterized by crossover rather close
to~$T_c$.\cite{fisher96}

While we acknowledge the potential value of the VT approach, we find, as will
be explained, that we cannot accept the validity of their analyses or of the
conclusions they draw.  Indeed, although Ising and classical behavior are
essentially different in the thermodynamic limit, they are far more difficult
to distinguish in the small systems that are accessible to numerical
simulations. Specifically, VT studied the heat capacity (i)~along the estimated
critical isochore (for $T>T_c$) and (ii)~along the anticipated critical
isotherm, for a wide range of densities.  In the first case, as mentioned, they
found no signs of a divergence in $C_V(T,\rho_c)$. In the second case, no
(finite-size rounded) peak was seen near the critical density.  It is this
latter observation that VT advance as strong evidence against Ising-type
critical behavior in the restricted primitive model, since, as they
illustrated, $C_V(T_c,\rho)$ in a Lennard-Jones fluid exhibits a clear,
system-size-dependent peak in the vicinity of $\rho=\rho_c$.  Here, we
reconsider this evidence, which stands unchallenged to date, either through new
simulations or via a reanalysis of the VT data.

Consider, first, the heat capacity along the critical isochore.  VT observe
that $C_V(T,\rho_c)$ increases almost linearly upon approach to~$T_c$ from
above, with no evidence of a divergence. They do, however, remark that this
might be due to the fact that all their observed temperatures lay within the
regime of finite-size rounding, where the correlation length is restricted by
the system size. We feel, rather, that it is the constraint $t \equiv
(T-T_c)/T_c > 0$ that might lead to premature conclusions. In Fig.~\ref{fig:mf}
we show the specific heat at constant volume for an infinite-range, van der
Waals or mean-field lattice gas (in which all particles interact equally) for a
number of system sizes.\cite{baxter82,thesis} The plots for density~$\rho =
\rho_c \equiv \frac{1}{2}\rho_{\rm max}$ represent the behavior along the
critical isochore, whereas the curves for $\rho = \frac{3}{4}\rho_c$ and $\rho
= \frac{5}{4}\rho_c$ (the system being symmetric around $\rho_c$) illustrate
the behavior along a noncritical isochore.  As expected, the peak heights are
lower if $\rho \neq \rho_c$; but the qualitative behavior of the specific heat
evidently persists even for relatively large deviations from the critical
isochore. (Thus, even moderately large errors in the estimate of~$\rho_c$ for
the RPM should not affect qualitative conclusions.)

The crucial point, however, is that (despite the absence of a divergence of
$C_V(T,\rho_c)$ in the thermodynamic limit) the mean-field plots display
pronounced size-dependent maxima for $T<T_c$.  Indeed, even though these peak
heights must saturate,\cite{thesis} whereas they diverge for an Ising-type
system with short-range interactions, the behavior of small systems is
qualitatively very similar in both cases.  In particular, the specific heats of
finite 3D Ising models and hard-core square-well fluids display maxima {\em
below\/}~$T_c$.\cite{orkoulas00a} Thus, it may be difficult to distinguish the
two types of behavior unless one has a sufficiently large range of system sizes
to allow extrapolation of the peak height and position.  Certainly, the linear
increase of $C_V(T,\rho)$ for $T \to T_c+$ for a given system size, as VT
observed for the RPM with $\rho \simeq \rho_c$, would seem to convey little
information regarding the nature of the critical behavior. This is basically a
consequence of the fact that for finite 3D systems the specific-heat maxima
seem invariably to occur {\em below\/} the true, limiting critical temperature.

In order to illustrate this point more concretely, we have carried out
high-resolution Monte Carlo simulations of a discretized version of the
RPM.\cite{panagiotopoulos99} This model differs from the continuum RPM only in
that the positions of the ions are restricted to lattice sites: the degree of
discretization is determined by the ratio, $\zeta$, of the ion
diameter~$\sigma$ to the lattice spacing~$a$. The continuum limit is recovered
by taking $\zeta \to \infty$.  It has been shown\cite{panagiotopoulos99} that
already for the small discretization parameters $\zeta=3$, $4$, and~$5$, this
model exhibits a liquid--vapor transition like the continuum RPM, with a
coexistence curve that approaches that of the continuum model very closely. We
have focused on $\zeta=5$, and carried out histogram-reweighted grand-canonical
simulations for simple cubic lattices of sizes up to $L=10\sigma$, which
corresponds to $(\zeta L/\sigma)^3 = 125\,000$ possible ion positions. Periodic
boundary conditions were employed.

\begin{figure}
\leavevmode
\centering
\epsfxsize \figurewidth
\epsfbox{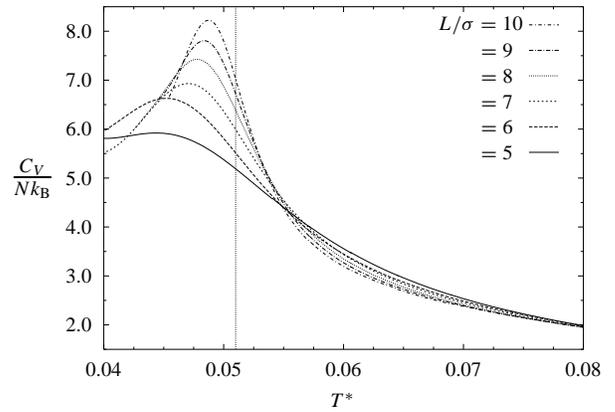}
\caption{The ionic specific heat, $C_V/Nk_{\rm B}$, of the RPM with
discretization parameter $\zeta=5$, along the estimated critical isochore.
Clear peaks rounded by finite size are evident below the estimated critical
temperature (vertical dashed line), although the behavior above $T_c$ conveys
significantly less information.}
\label{fig:cv_isochore}
\end{figure}

The strong ion pairing at low temperatures makes grand-canonical simulations
especially time consuming. However, we view canonical simulations as inherently
dangerous owing to the important role of density fluctuations in the vicinity
of the critical point. (See also further comments below.)  A detailed study of
these and related data is in progress,\cite{luijten00c} but {\em preliminary\/}
examination suggests a reduced critical temperature $T_c^* \simeq 0.051$ and a
critical density $\rho_c^* \simeq 0.068$. (See, e.g.,
Refs.~\onlinecite{fisher94,fisher96,fisher93,moreira99,caillol94,orkoulas94}
for the standard definitions of reduced units for the RPM.)  Our estimates
agree with the suggestion of Ref.~\onlinecite{panagiotopoulos99} that the RPM
with $\zeta=5$ has a $T_c^*$ slightly higher than that of the continuum RPM
($T_c^* \simeq 0.049$\cite{orkoulas94,caillol96,yan99}), although the estimate
of VT is $T_c^* \simeq 0.052$.  Figure~\ref{fig:cv_isochore} shows the specific
heat of this model for $\rho \simeq \rho_c$ over a relatively wide temperature
range around~$T_c$. As in Fig.~\ref{fig:mf}, system-size dependent maxima are
indeed observed at temperatures that clearly approach~$T_c$ from below. Their
variation with $L$ suggests quite strongly that Ising-type behavior cannot be
ruled out.

On the other hand, VT's results for the density dependence of~$C_V$
at $T=T_c$ present, at first sight, a more challenging puzzle.  Indeed, their
data appeared to reveal a remarkable difference between the RPM and a
Lennard-Jones fluid. For the former, $C_V/Nk_{\rm B}$ decreased monotonically
with increasing density, without any evident marked dependence on system size.
%(The results for VT's smallest system size deviate slightly from the remaining
%data; but this seems to result from the fact that they were obtained at a
%different temperature, rather than reflecting finite-size effects.) 
The corresponding curves VT present for a Lennard-Jones fluid, on the contrary,
exhibit a pronounced peak in the vicinity of the critical density~$\rho_c$
that, furthermore, increases with system size.

\begin{figure}
\leavevmode
\centering
\epsfxsize \figurewidth
\epsfbox{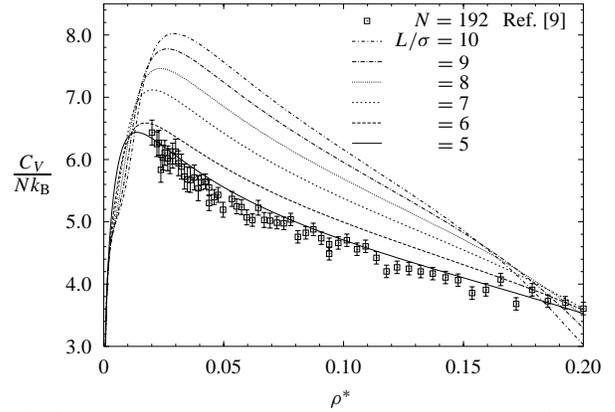}
\caption{The constant-volume specific heat on the estimated critical isotherm
of the discretized RPM (for $\zeta = 5$), compared with the corresponding VT
data (Ref.~\protect\onlinecite{valleau98}).}
\label{fig:cv_isotherm1}
\end{figure}

Why is a peak for the RPM apparently absent?  Without doubt, any maximum or
divergence related to criticality in the RPM will be rounded and shifted in
finite systems, irrespective of the actual universality class. Furthermore, one
must be prepared for strong finite-size effects that are likely to be distorted
relative to LJ-type model fluids in light of the shape of the RPM coexistence
curve, which is highly asymmetric.\cite{fisher94,fisher96,fisher93} This
expectation is indeed confirmed by Fig.~\ref{fig:cv_isotherm1}, where the
continuous plots show our simulation data for the specific heat of the
discretized RPM for six different system sizes at $T^* =0.051 \simeq T_c^*$.
For each system size there is a clear maximum, at a density that appears to
approach the critical density from below.  Unfortunately, however, the data of
VT (shown for their system size $N=192$) did not extend to sufficiently low
densities to cover these peaks. Also noteworthy is that their data actually
agree rather well with ours for the discretized RPM, at a system size between
$L=5\sigma$ and $L=6\sigma$, even though this is considerably smaller than the
dimension $\tilde{L}_N \equiv (N/\rho)^{1/3} \simeq 13\sigma$ suggested by the
particle number~$N=192$. This difference might be due to the $\zeta=5$
discretization; but if $T_c$ really is lower for the
RPM,\cite{panagiotopoulos99} VT's data were taken at an isotherm somewhat
above~$T_c$.  On the other hand, one cannot exclude the possibility that the
difference reflects a relative limitation of the {\em canonical\/} simulations
performed by VT\@.  The fixed particle number may, in effect, suppress the
characteristic density fluctuations, with a consequential relative suppression
and enhanced rounding of various maxima in finite systems.  This might also
explain why VT observe negligible changes in $C_V$ when increasing the number
of ions by 50\% from 128 to~192.  However, the precise nature of the
finite-size effects themselves is currently unclear in the framework of the
$T$-and-$\rho$-scaling Monte Carlo used by VT.

Now consider the qualitative differences observed by VT between the specific
heats of the RPM and the LJ fluid. These differences also prove to be a
consequence of the low critical density of the RPM, compared to $\rho_c$ for
simple fluids; they do {\em not\/} reflect any significant possible difference
in the nature of the critical behavior of the two systems.  To see this, recall
that in comparing the {\em configurational\/} heat capacity at constant volume
for different fluids, or for a fluid and a lattice model, the more basic
quantity for criticality and phase separation is the heat capacity {\em per
unit volume\/} (or $C_V$ {\em density\/}) rather than the specific heat (or
heat capacity per {\em particle\/}).\cite{fisher64} In addition to the
arguments presented in Ref.~\onlinecite{fisher64}, namely the greater
naturalness of the grand-canonical ensemble and the expectation that {\em
spatial\/} fluctuations most directly characterize critical divergences, note
that the field-theoretic viewpoint of critical phenomena and the
renormalization-group approach\cite{fisher-rgreview} bear out the conclusion
that the number of degrees of freedom {\em per unit volume\/} plays the most
fundamental role.

\begin{figure}
\leavevmode
\centering
\epsfxsize \figurewidth
\epsfbox{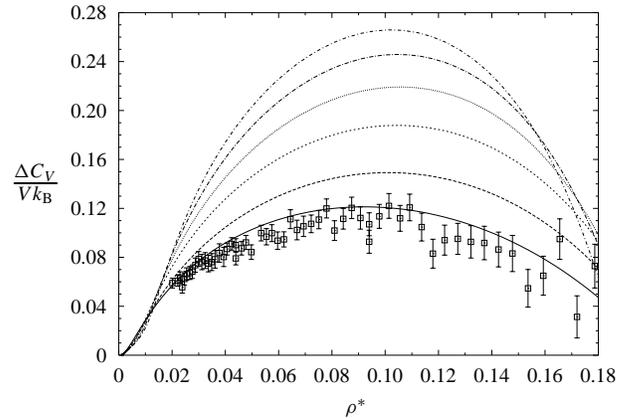}
\caption{As in Fig.~\protect\ref{fig:cv_isotherm1}, but now the heat-capacity
{\em density\/} is plotted and a linear ``background term'' $C_V^0/Vk_{\rm B} =
3.5\rho^*$ has been subtracted.}
\label{fig:cv_isotherm2}
\end{figure}

\begin{figure}
\leavevmode
\centering
\epsfxsize \figurewidth
\epsfbox{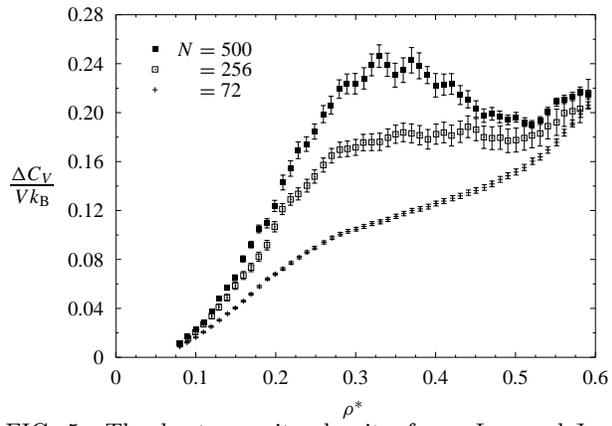}
\caption{The heat-capacity density for a Lennard-Jones fluid, as
derived from the data of Valleau and Torrie
(Ref.~\protect\onlinecite{valleau98}). As in
Fig.~\protect\ref{fig:cv_isotherm2}, a ``background'' $C_V^0/Vk_{\rm B} =
1.6\rho^*$ has been subtracted from $C_V/Vk_{\rm B}$.}
\label{fig:cv_lj}
\end{figure}

Thus, in Fig.~\ref{fig:cv_isotherm2} we have plotted the heat-capacity density,
$C_V/Vk_{\rm B}$, for the RPM as a function of $\rho$ for $T^* = 0.051$, with a
``harmless background'' proportional to $\rho$ subtracted.  Instead of a
monotonically decreasing function, we now obtain a net ``energy fluctuation''
that displays a clear maximum as a function of~$\rho$, of a height that
increases systematically with size.  If we replot VT's data for the
Lennard-Jones fluid in the analogous way, as in Fig.~\ref{fig:cv_lj}, we find
similar behavior: indeed, the subtracted heat-capacity density of the LJ fluid
on the critical isotherm exhibits clear finite-size maxima as a function of the
density, with a systematic size dependence that, as for the RPM, suggests a
monotonic increase (as $L$ or $N \to \infty$) peaking in the vicinity of the
critical density.

In summary, the constant-volume heat capacity is a useful quantity in the study
of the critical behavior of the restricted primitive model for ionic fluids.
However, as we have demonstrated, care must be exercised before concluding that
a maximum or a critical divergence is absent. Along both the critical isochore
and the critical isotherm, numerical results for finite systems exhibit clear
maxima, contrary to the suggestions of Ref.~\onlinecite{valleau98}.
Furthermore, inasfar as one observes an overall qualitative difference between
the specific heat for the RPM and a fluid with Lennard-Jones interactions, the
effect is primarily due to the large difference in critical densities. It has
no relevance to possible differences in critical behavior---for which our
current data allow few definitive statements.  Conversely, if one considers the
heat-capacity {\em density}, the two systems exhibit qualitatively rather
similar behavior. Whether, ultimately, that reflects the same or distinct
critical universality classes remains to be determined on the basis of
fluctuations not only of the energy, as observed in the specific heat, but also
of fluctuations in density.\cite{orkoulas00a,luijten00c}

\acknowledgements

We thank Dr.\ G. Orkoulas for informative discussions.  The support of the
National Science Foundation (through Grant No.\ CHE 99-81772 to M.E.F.)
and of the Department of Energy, Office of Basic Energy Sciences (through Grant
No.\ DE-FG02-98ER14858 to A.Z.P.) is gratefully acknowledged.

%\bibliographystyle{prsty}
%\bibliography{journals,electrolyte,misc}

\raggedcolumns

\end{multicols*}

\end{document}